\documentclass[a4paper,fleqn]{cas-dc}

\usepackage[numbers]{natbib}

\def\tsc#1{\csdef{#1}{\textsc{\lowercase{#1}}\xspace}}
\tsc{WGM}
\tsc{QE}
\tsc{EP}
\tsc{PMS}
\tsc{BEC}
\tsc{DE}

\begin{document}
\let\WriteBookmarks\relax
\def\floatpagepagefraction{1}
\def\textpagefraction{.001}

\shorttitle{Spatial scaling of heterogeneity transfer from boron concentration to absorbed dose in boron neutron capture therapy}
\shortauthors{R.~Muslim, B.~Pidhatika, S.~Sutanto, R.~Fardela, R.~N.~Wijaya }

\title[mode = title]{Spatial scaling of heterogeneity transfer from boron concentration to absorbed dose in boron neutron capture therapy}    
\author[1,2, 3]{Roni Muslim}[orcid= 0000-0001-6925-5923]
\ead{roni.muslim@apctp.org}
\cormark[1]
\cortext[cor1]{Corresponding author}
\author[3]{Bidhari Pidhatika}
\ead{bidhari.pidhatika@brin.go.id}

\author[3]{Sutanto Sutanto}
\ead{suta012@brin.go.id}

\author[4]{Ramacos Fardela}[orcid=0000-0002-6991-4156]
\ead{ramacosfardela@sci.unand.ac.id}

\author[3]{Rio Natanel Wijaya}
\ead{rio.natanel.wijaya@brin.go.id}

\affiliation[1]{
    organization={Asia Pacific Center for Theoretical Physics (APCTP)},
    city={Pohang},
    postcode={37673},
    country={Republic of Korea}
}
    
\affiliation[2]{organization={Research Center for Quantum Physics,
    National Research and Innovation Agency (BRIN)},
    city={South Tangerang},
    postcode={15314}, 
    country={Indonesia}}
    
\affiliation[3]{organization={Indonesian Nuclear Technology Polytechnic,
    National Research and Innovation Agency (BRIN)},
    city={Yogyakarta},
    postcode={55281}, 
    country={Indonesia}}

\affiliation[4]{
    organization={Department of Physics, Faculty of Mathematics and Natural Science, Universitas Andalas},
    state={Sumatera Barat},
    postcode={25163},
    country={Indonesia}
}

\begin{abstract}
We investigate how the amplitude and spatial correlation length of boron-concentration fluctuations affect the boron-induced absorbed-dose distribution in boron neutron capture therapy (BNCT). The spatial energy-deposition response of alpha particles and lithium nuclei is obtained using Geant4 and convolved with three-dimensional correlated lognormal boron fields under a uniform neutron field. Approximately $90\%$ of the reaction-product energy is deposited within a radius of $6.00~\mu\mathrm{m}$, with a characteristic deposition length of $\ell_B\approx2.05~\mu\mathrm{m}$. Dose heterogeneity increases with both the fluctuation amplitude $H$ and correlation length $\xi$, while the transfer factor $S=CV_D/H$ exhibits an approximate scaling collapse as a function of $\xi/\ell_B$. Analytical Gaussian and lognormal--Gaussian models provide reference predictions for this behavior. At short correlation lengths, the Geant4 response yields a higher transfer factor than a Gaussian response with the same characteristic length, highlighting the influence of the energy-deposition shape on dose smoothing. For $D_{95}$, defined as the dose received by at least $95\%$ of the modeled volume, the relative difference between homogeneous and heterogeneous boron-dose models reaches $(86.62\pm1.07)\%$ at $H=1.2$ and a target ratio $\xi/\ell_B=6$. We show that mean boron concentration alone is insufficient to characterize spatial nonuniformity of the boron-dose component.
\end{abstract}




\maketitle

\section{Introduction}
\label{sec:introduction}

Boron neutron capture therapy (BNCT) is a
radiotherapeutic approach that combines the
delivery of compounds containing
$^{10}\mathrm{B}$ with neutron irradiation.
Following neutron capture, $^{10}\mathrm{B}$
produces alpha particles and $^{7}\mathrm{Li}$
nuclei that deposit their energy over distances
of only a few micrometers. These short ranges
allow energy deposition to occur close to the
boron capture sites, making the spatial
distribution of boron an important factor
in determining the absorbed dose
\cite{barth2005bnct,barth2012status,dymova2020status}.
The development of accelerator-based neutron
sources has expanded the clinical application
of BNCT, particularly for certain head and
neck cancers, while investigations involving
brain tumors and other malignancies continue
to advance
\cite{hirose2021jhn002,shen2024clinician,
younous2024systematic,zhou2024clinical,
barth2024status,zhou2025clinical,
kutuk2026cns,zheng2026renaissance}.

The effectiveness of BNCT depends not only
on the characteristics of the neutron beam
but also on the ability of boron delivery
agents to reach the target tissue and
achieve a suitable distribution of
$^{10}\mathrm{B}$. Compounds such as
boronophenylalanine (BPA) and sodium
borocaptate (BSH) have long been investigated
for this purpose, while the development of
targeted molecules, polymers, and
nanoparticle-based carriers aims to improve
boron selectivity and retention
\cite{montihu2023optimizing,barth2024agents,
xu2024agents,li2025nanomaterials,
murilla2025agents,yang2026agents,
tang2026carriers,selg2026crossroads,
yadav2026nanocarriers,xie2026nanoparticles}.
Nevertheless, boron concentration is not
necessarily uniform, either across different
regions of a tumor or at the cellular and
subcellular levels. Measuring and mapping
boron distributions with sufficient spatial
detail remains challenging. The mean boron concentration alone may not
adequately represent the spatial variations
that govern energy deposition
\cite{zhou2025heterogeneity,ma2026clinical,
selg2026quantification}.

The importance of boron distribution at
small spatial scales has been demonstrated
through microdosimetric studies.
Sato \textit{et al.} developed a model
that accounts for intra- and intercellular
variations in $^{10}\mathrm{B}$ distribution,
showing that microscopic dose distributions
are important for interpreting the biological
effects of BNCT
\cite{sato2018microdosimetry}.
Subsequent studies have further developed
stochastic modeling approaches, particle
transport calculations, and methods for
estimating biological responses at the
cellular scale
\cite{sato2021smk,furuta2021phits,
shigehira2026lismec,yamazaki2026survival}.
Recent microdosimetric investigations and
measurements have also emphasized that
energy deposition at micrometer scales
cannot always be adequately characterized
by the mean absorbed dose alone
\cite{selva2024microdosimetry,
santacruz2026microdosimetry,
marcaccio2026microdistribution}.
These findings highlight the relevance
of boron heterogeneity to BNCT dosimetry,
although the spatial distribution of
absorbed dose should be distinguished
from direct predictions of biological
response.

Monte Carlo (MC) methods are widely used in
BNCT treatment planning to calculate
the contributions from different radiation
components. Advances in computational
tools and image-based modeling have
improved the evaluation of absorbed-dose
distributions
\cite{zhong2023compass,pistone2025gate,
zhu2026tps,kakino2026wholebody}.
However, some dosimetric approaches
still assume uniform boron concentrations
or uptake ratios within specified regions.
Teng \textit{et al.} demonstrated that
accounting for heterogeneous boron
distributions can alter calculated
tumor doses
\cite{teng2023heterogeneity}.
Subsequent studies using
$^{18}\mathrm{F}$-BPA positron emission
tomography (PET) have also reported
differences in dosimetric indices when
voxel-level boron variations are considered,
while highlighting limitations associated
with image resolution and segmentation
\cite{kobayashi2025pet,
teng2026hyposegmentation,teng2026mucosa}.
At the same time, BNCT dose evaluation
must account for the neutron field and
radiation components other than those
arising from boron neutron capture
\cite{narita2024internal,hu2025monitoring}.

These studies establish the importance
of boron heterogeneity but leave a
more fundamental physical question:
how does the spatial scale of boron
concentration fluctuations determine
the degree of dose smoothing caused
by alpha-particle and lithium-nucleus
transport? Two boron distributions
may have the same mean concentration
and coefficient of variation while
differing in their spatial correlation
lengths, which characterize the typical
sizes of boron-rich and boron-poor
regions. Because the reaction products
deposit energy over finite distances,
small-scale boron fluctuations are
expected to undergo stronger spatial
smoothing than larger structures.
This behavior cannot be determined
from the mean concentration or the
amplitude of boron heterogeneity
alone without considering the spatial
characteristics of energy deposition.
A framework that independently controls
the heterogeneity amplitude, boron
correlation length, and energy-deposition
response is therefore needed.

In this study, these effects are
investigated using correlated lognormal
boron concentration fields and a spatial
energy-deposition response obtained
from Geant4 simulations
\cite{agostinelli2003geant4,
allison2016geant4}.
The boron-dose component is calculated
through three-dimensional convolution
in a homogeneous medium under the
assumption of a uniform neutron field.
To quantify how variations in boron
concentration are transferred to the
absorbed-dose distribution, we define
a heterogeneity transfer factor,
$S=CV_D/H$, where $H$ is the coefficient
of variation of boron concentration
and $CV_D$ is the corresponding
coefficient of variation of absorbed
dose. The dependence of $S$ on the
ratio $\xi/\ell_B$, where $\xi$
is the boron correlation length
and $\ell_B$ is the characteristic
energy-deposition length, is then
compared with an ideal Gaussian
prediction. By applying the Geant4
and Gaussian responses to identical
boron concentration fields, the
influence of the spatial shape of
energy deposition can be examined
directly.

Finally, the dosimetric consequences
are evaluated using $D_{95}$,
defined as the dose received by at
least $95\%$ of the modeled volume.
This quantity is compared with the
corresponding result obtained under
the assumption of a homogeneous boron
distribution with the same mean
concentration. The present study
focuses on the relationship between
the spatial structure of boron
concentration and the absorbed-dose
component arising from boron neutron
capture, rather than on complete
clinical dose calculations or the
biological effectiveness of BNCT.
Within this framework, we aim to
identify the conditions under which
the homogeneous-boron approximation
may underestimate dose nonuniformity
and to clarify how spatial correlation
and the shape of the energy-deposition
response jointly influence the
resulting dose distribution.

\section{Model and Methods}
\label{sec:model_methods}

\subsection{Spatial model of boron distribution}
\label{sec:boron_distribution}

This study employs a computational approach
to investigate how spatial heterogeneity
in boron concentration influences energy
deposition in boron neutron capture therapy
(BNCT). To isolate this effect, tissue is
represented by a homogeneous medium with
fixed density and composition, while the
$^{10}\mathrm{B}$ concentration is allowed
to vary spatially. The neutron field is
assumed to be uniform throughout the
computational domain, so variations in
absorbed dose arise from the spatial
distribution of boron and the transport
of reaction-product energy. The model
is not intended to reproduce patient-specific
anatomy but rather to examine how the
amplitude and spatial scale of boron
heterogeneity influence the resulting
dose nonuniformity.

The boron concentration is represented
by a three-dimensional scalar field,
$C_B(\mathbf{r})$, with mean value
$\bar C_B$. To generate heterogeneous
distributions with strictly positive
concentrations, a correlated lognormal
random field is used:
\begin{equation}
C_B(\mathbf{r})
=
\bar C_B
\exp\left[
sX(\mathbf{r})-\frac{s^2}{2}
\right].
\label{eq:boron_field}
\end{equation}
Here, $X(\mathbf{r})$ is a stationary
Gaussian random field with zero mean
and unit variance. Its spatial
correlation function is given by
\begin{equation}
\left\langle
X(\mathbf{r})X(\mathbf{r}')
\right\rangle
=
\exp\left[
-\frac{|\mathbf{r}-\mathbf{r}'|^2}{2a^2}
\right],
\label{eq:spatial_correlation}
\end{equation}
where $a$ denotes the correlation length
of the underlying Gaussian field.
The degree of boron heterogeneity is
characterized by the coefficient of
variation,
\begin{equation}
H
=
\frac{\sigma_B}{\bar C_B},
\label{eq:boron_heterogeneity}
\end{equation}
where $\sigma_B$ is the standard
deviation of boron concentration.
The lognormal parameter is determined
by $s=\sqrt{\ln(1+H^2)}$, ensuring that
the prescribed mean concentration and
heterogeneity amplitude are independently
controlled.

The spatial correlation of the boron
concentration field differs from that
of the underlying Gaussian field because
of the lognormal transformation. In
particular, the normalized autocorrelation
function of $C_B(\mathbf{r})$ can be
expressed analytically as
\begin{equation}
\rho_B(r)
=
\frac{
\exp\left[
s^2\exp\left(-r^2/2a^2\right)
\right]-1
}{
\exp(s^2)-1
}.
\label{eq:boron_autocorrelation}
\end{equation}
We define the boron correlation length
$\xi$ as the distance at which
$\rho_B(\xi)=\exp(-1/2)$. This definition
distinguishes the spatial scale of the
boron concentration fluctuations from
the correlation length $a$ used to
generate the Gaussian field. For a
prescribed heterogeneity amplitude
$H>0$ and target correlation length
$\xi$, the corresponding Gaussian
correlation length is therefore
\begin{equation}
a
=
\frac{\xi}{
\sqrt{
2\ln\left[
\frac{\ln(1+H^2)}
{\ln(1+H^2e^{-1/2})}
\right]
}
}.
\label{eq:correlation_length_mapping}
\end{equation}
This relation accounts for the change
in spatial correlation introduced by
the lognormal transformation and allows
the prescribed heterogeneity amplitude
and boron correlation length to be
varied independently. In the limit
of weak heterogeneity, $a$ approaches
$\xi$, whereas their difference becomes
more pronounced as $H$ increases.

In the numerical analysis, the prescribed
mean boron concentration is kept constant,
while $H$ and $\xi$ are varied
systematically. For each parameter
combination, the Gaussian correlation
length is determined from
Eq.~\eqref{eq:correlation_length_mapping},
and independent random realizations
of the boron concentration field are
generated. The actual correlation length
of each realization is then evaluated
from its spatial autocorrelation function
to account for finite-domain fluctuations
and numerical discretization. Two boron distributions can have the
same mean concentration and heterogeneity
amplitude but different spatial extents
of boron-rich and boron-poor regions.
By independently controlling these
properties and averaging over multiple
realizations, the effects of heterogeneity
amplitude and correlation length on the
absorbed-dose distribution can be
examined separately.

\subsection{Particle transport and spatial
energy-deposition response}
\label{sec:energy_kernel}

The transport of charged particles
produced by neutron capture in
$^{10}\mathrm{B}$ is simulated using
the MC toolkit Geant4
\cite{agostinelli2003geant4,allison2016geant4}.
The reaction considered is
\begin{equation}
{}^{10}\mathrm{B}+n
\longrightarrow
{}^{7}\mathrm{Li}+\alpha,
\label{eq:boron_capture}
\end{equation}
which produces charged particles
with ranges of only a few micrometers
in tissue. In the dominant reaction
branch, with a probability of
approximately $93.7\%$, the alpha
particle and lithium nucleus have
kinetic energies of $1.47$ and
$0.84$ MeV, respectively. The
remaining branch, with a probability
of approximately $6.3\%$, produces
an alpha particle and a lithium
nucleus with kinetic energies of
approximately $1.78$ and $1.01$ MeV,
respectively. Both branches are
included according to their
respective probabilities. Previous
MC track-structure studies
have also examined the influence
of lithium-ion transport and
interaction cross sections on
BNCT microdosimetry
\cite{han2023microdosimetry}.
The present study considers only
energy deposition from alpha
particles and lithium nuclei,
excluding gamma-ray contributions
and other neutron-induced
interactions.

Particle transport is simulated
in water as a simplified
representation of biological tissue.
Reaction products are generated
at a single point at the center
of the medium, with emission
directions sampled isotropically.
Each particle species and reaction
branch is simulated separately,
and the deposited energy is recorded
as a function of distance from
the reaction site using concentric
spherical shells. The radial
contributions are combined according
to the corresponding branching
probabilities to construct the
normalized spatial energy-deposition
response, $K_B(r)$. This response
describes the average spatial
distribution of energy deposited
by the charged reaction products
around a neutron-capture site.

\begin{figure*}[pos=tb]
    \centering
    \includegraphics[width=0.8\linewidth]
    {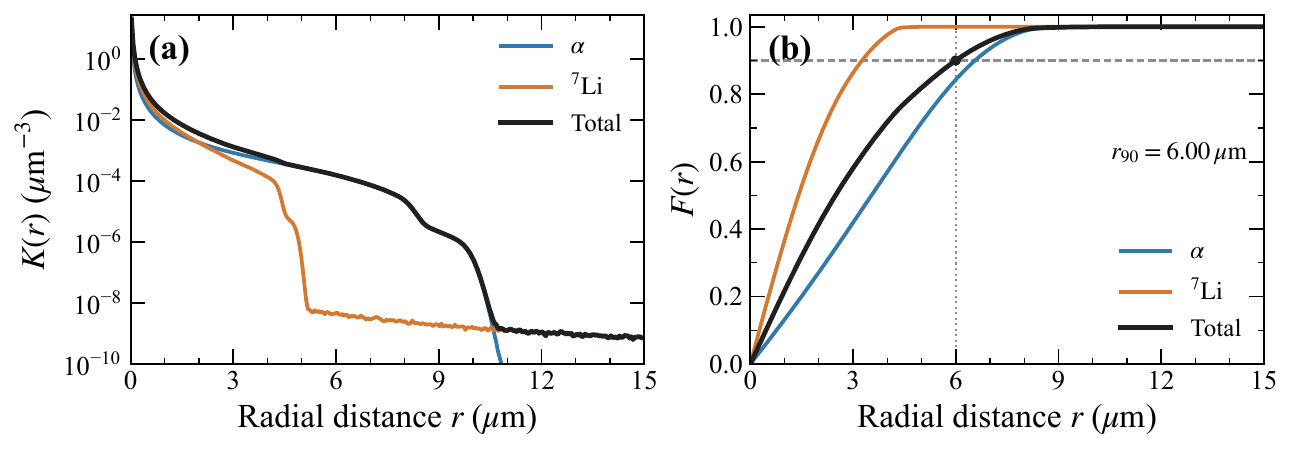}
    \caption{Spatial distribution of energy
    deposited by alpha particles ($\alpha$)
    and lithium nuclei ($^{7}\mathrm{Li}$)
    produced in the
    $^{10}\mathrm{B}(n,\alpha)^{7}\mathrm{Li}$
    reaction, obtained from Geant4
    MC simulations in water.
    (a) Normalized radial energy-deposition
    responses for the individual particle
    species and their combined contribution.
    (b) Cumulative fraction of deposited
    energy as a function of distance
    from the reaction site. The dashed
    horizontal line indicates the
    $90\%$ energy fraction, while the
    dotted vertical line marks
    $r_{90}\approx6.00~\mu\mathrm{m}$.
    The results show that energy deposition
    is localized, with alpha particles
    exhibiting longer ranges than
    lithium nuclei.}
    \label{fig:bnct_kernel}
\end{figure*}

Figure~\ref{fig:bnct_kernel}
illustrates the spatial
energy-deposition response obtained
from the Geant4 simulations.
In Fig.~\ref{fig:bnct_kernel}(a),
energy deposition from
$^{7}\mathrm{Li}$ nuclei is more
concentrated near the reaction
site, whereas alpha particles
deposit energy over longer
distances. This difference is
also evident in the cumulative
energy distributions shown in
Fig.~\ref{fig:bnct_kernel}(b),
where approximately $90\%$
of the total energy is deposited
within a radius of
$r_{90}\approx6.00~\mu\mathrm{m}$.
These results demonstrate that
the reaction products deposit
energy over micrometer-scale
distances, suggesting that the
spatial extent of boron concentration
variations is important in
determining the resulting
dose nonuniformity.

To relate boron concentration to
the absorbed-dose distribution,
the neutron field is assumed to
be spatially uniform, and changes
in boron concentration are assumed
not to significantly perturb the
neutron distribution. Under these
conditions, the local neutron-capture
rate is proportional to
$C_B(\mathbf{r})$. Because the
medium is homogeneous, the
energy-deposition response depends
only on the relative displacement
between the reaction site and
the energy-deposition location.
The absorbed-dose distribution
arising from the boron component
can therefore be expressed as
a convolution,
\begin{equation}
D_B(\mathbf{r})
=
A
\int
K_B(\mathbf{r}-\mathbf{r}')
C_B(\mathbf{r}')
\,d^3\mathbf{r}',
\label{eq:dose_convolution}
\end{equation}
where $A$ is a normalization
factor incorporating neutron
fluence, capture probability,
mean reaction-product energy,
and conversion to absorbed dose.
This formulation allows dose
distributions for different boron
configurations to be calculated
through numerical convolution
without repeating the particle
transport simulations for each
realization. Its accuracy could
be further assessed by comparison
with direct particle transport
simulations for selected
heterogeneous configurations.

\subsection{Analytical theory and numerical analysis}
\label{sec:scaling_theory}

The relationship between boron heterogeneity
and absorbed dose is examined by separating
each field into its mean and fluctuating
components. Based on the convolution model
in Eq.~\eqref{eq:dose_convolution}, the Fourier
transforms of the dose and boron concentration
fluctuations satisfy
\begin{equation}
\widetilde{\delta D}_B(\mathbf{k})
=
A\widetilde{K}_B(\mathbf{k})
\widetilde{\delta C}_B(\mathbf{k}).
\label{eq:fourier_dose}
\end{equation}
This relation shows that particle transport
acts as a spatial filter on boron concentration
fluctuations. Variations occurring on spatial
scales smaller than the energy-deposition
range tend to be smoothed, whereas larger-scale
variations are more readily preserved in the
absorbed-dose distribution. Defining $S_B(k)$
and $S_D(k)$ as the power spectra of boron
concentration and dose fluctuations,
respectively, gives
\begin{equation}
S_D(k)
=
A^2
\left|
\widetilde{K}_B(k)
\right|^2
S_B(k).
\label{eq:dose_power_spectrum}
\end{equation}
This spectral relation connects the spatial
structure of boron concentration with the
resulting absorbed-dose fluctuations, with
the energy-deposition response determining
how strongly fluctuations at different
spatial scales are attenuated.

To quantify the transfer of boron heterogeneity
to the absorbed-dose distribution, we introduce
the heterogeneity transfer factor
\begin{equation}
S
=
\frac{CV_D}{H},
\qquad
CV_D
=
\frac{\sigma_D}{\langle D_B\rangle},
\label{eq:transfer_factor}
\end{equation}
where $\sigma_D$ is the standard deviation
of absorbed dose. A small value of $S$
indicates that particle transport substantially
smooths spatial variations in boron
concentration, whereas $S$ approaching unity
indicates that dose heterogeneity closely
follows the underlying boron distribution.
The transfer factor is defined for $H>0$,
with the homogeneous boron distribution
serving as the reference condition.
For a normalized energy-deposition response,
and neglecting finite-domain effects,
the transfer factor can be expressed as
\begin{equation}
S^2
=
\frac{
\displaystyle
\int
S_B(k)
\left|
\widetilde{K}_B(k)
\right|^2
\,d^3\mathbf{k}
}{
\displaystyle
\int
S_B(k)\,d^3\mathbf{k}
}.
\label{eq:transfer_spectrum}
\end{equation}
Equation~\eqref{eq:transfer_spectrum}
shows that heterogeneity smoothing depends
on both the spatial spectrum of boron
concentration and the shape of the
energy-deposition response. The transfer of heterogeneity cannot
generally be characterized by the fluctuation
amplitude or a single correlation length
without specifying the underlying spatial
correlation structure.

To obtain a simple analytical prediction,
we first consider a reference model with
a Gaussian boron concentration covariance
and a Gaussian energy-deposition response
characterized by a length scale $\ell_B$.
In three dimensions, integration of
Eq.~\eqref{eq:transfer_spectrum} yields
\begin{equation}
S_G
=
\left(
\frac{\xi^2}
{\xi^2+2\ell_B^2}
\right)^{3/4}.
\label{eq:analytical_scaling}
\end{equation}
This expression predicts that heterogeneity
transfer is governed by the ratio
$\xi/\ell_B$, which compares the boron
correlation length with the characteristic
energy-deposition length. When
$\xi\ll\ell_B$, small-scale boron
fluctuations undergo substantial smoothing,
whereas for $\xi\gg\ell_B$, energy deposition
becomes relatively localized within the
boron structures and $S_G$ approaches
unity. Equation~\eqref{eq:analytical_scaling}
is exact for the assumed Gaussian covariance
and response, but it does not fully describe
the correlated lognormal boron fields
used in the numerical calculations.

To account for the lognormal structure
of the boron concentration field, we extend
the analytical treatment while retaining
the Gaussian energy-deposition response.
Using the spatial correlation model defined
in Sec.~\ref{sec:boron_distribution}, the
resulting heterogeneity transfer factor
can be expressed as
\begin{equation}
S_{\mathrm{LG}}^2
=
\frac{1}{e^{s^2}-1}
\sum_{m=1}^{\infty}
\frac{s^{2m}}{m!}
\left(
\frac{a^2}
{a^2+2m\ell_B^2}
\right)^{3/2},
\label{eq:lognormal_gaussian_scaling}
\end{equation}
where $s^2=\ln(1+H^2)$ and $a$ is the
correlation length of the underlying
Gaussian field. As established in
Sec.~\ref{sec:boron_distribution},
$a$ is related to the boron correlation
length $\xi$ through the lognormal
transformation. Unlike the Gaussian
reference prediction, this expression
retains the dependence on heterogeneity
amplitude through the spatial covariance
of the lognormal field. In the limit
of weak heterogeneity, $a$ approaches
$\xi$, and Eq.~\eqref{eq:lognormal_gaussian_scaling}
reduces to Eq.~\eqref{eq:analytical_scaling}.
The two analytical expressions therefore
provide complementary descriptions of
heterogeneity transfer. The Gaussian
reference model describes the dependence
on the correlation-length ratio for an
ideal Gaussian covariance, while the
lognormal--Gaussian model additionally
accounts for changes in the correlation
structure associated with the lognormal
transformation. Their derivations are
presented in
Appendix~\ref{app:gaussian_scaling}.

In the numerical calculations, the
characteristic energy-deposition length
is determined from the second moment
of the spatial response,
$\ell_B^2=\langle r^2\rangle_K/3$.
The Geant4 simulations, performed
with a radial resolution of
$0.05~\mu\mathrm{m}$, yield
$\ell_B\approx2.05~\mu\mathrm{m}$.
The actual boron correlation length,
$\xi$, is determined from the normalized
autocorrelation function of the boron
concentration field as the distance
at which the correlation decreases
to $\exp(-1/2)$. This measured value
is distinguished from the target
correlation length specified during
field generation to account for
finite-domain fluctuations and numerical
discretization. Absorbed-dose distributions
are calculated through convolution using
three-dimensional fast Fourier transforms
with periodic boundary conditions.
Spatial visualizations use a $512^3$
grid with a domain length of
$64~\mu\mathrm{m}$ along each direction,
whereas quantitative analyses use
a $512^3$ grid with a domain length
of $128~\mu\mathrm{m}$. Different
combinations of heterogeneity amplitude
$H$ and correlation-length ratio
$\xi/\ell_B$ are examined to characterize
their effects on the resulting
dose distribution.

To assess the influence of the spatial
shape of energy deposition, additional
calculations are performed using a
Gaussian response with the same
characteristic length $\ell_B$ as
the Geant4 response. Both responses
are applied to identical realizations
of the lognormal boron concentration
field, allowing their effects on the
transfer factor to be compared directly.
The analytical predictions in
Eqs.~\eqref{eq:analytical_scaling}
and \eqref{eq:lognormal_gaussian_scaling}
provide reference results for examining
the influence of boron correlation
structure and energy-deposition response.
In particular, the lognormal--Gaussian
prediction provides an analytical
benchmark for calculations using
the Gaussian response, while the
comparison between the numerical
Gaussian and Geant4 responses isolates
the contribution of the energy-deposition
kernel shape. For each parameter
combination in the main quantitative
analyses, 10 independent random
realizations are generated. Results
are reported as ensemble averages,
with statistical uncertainties expressed
as standard errors of the mean.

The dosimetric consequences of boron
heterogeneity are further evaluated
using $D_{95}$, defined as the fifth
percentile of the absorbed-dose
distribution, corresponding to the
dose received by at least $95\%$
of the modeled volume. To enable
comparisons across parameter
combinations, the dose distribution
is normalized by its mean,
$d_B(\mathbf{r})=
D_B(\mathbf{r})/\langle D_B\rangle$.
Under this normalization, a homogeneous
boron distribution produces a relative
dose of unity throughout the domain.
The difference in $D_{95}$ between
homogeneous and heterogeneous boron
models is therefore used to assess
the effect of assuming a uniform
boron concentration on the coverage
of the boron-dose component.
Unlike the transfer factor, which
characterizes the overall amplitude
of dose fluctuations, $D_{95}$ is
sensitive to the lower tail of the
dose distribution and can therefore
reveal differences that are not
fully captured by the coefficient
of variation alone.

The numerical stability of the
three-dimensional dose calculation
is assessed separately from the
convergence of the Geant4
energy-deposition response. Grid
refinement is examined at a fixed
physical domain size, whereas
finite-domain sensitivity is
evaluated by increasing the domain
size while keeping the spatial
grid spacing unchanged. The tests
focus on $H=1.2$ and a target
correlation-length ratio
$\xi/\ell_B=6$, corresponding
to the largest relative difference
in $D_{95}$ within the investigated
parameter range. The reference
configuration of $512^3$ grid
points and a domain length of
$128~\mu\mathrm{m}$ is compared
with a $640^3$ grid at the same
domain length and with a
$640^3$ grid in a
$160~\mu\mathrm{m}$ domain.
For each configuration, 20
independent random realizations
are generated, and the transfer
factor $S$ and relative difference
in $D_{95}$ are evaluated to
assess their sensitivity to the
numerical settings. These
comparisons provide a measure
of numerical robustness under
grid refinement and domain
enlargement, while allowing
statistical fluctuations between
independent realizations to be
taken into account. The numerical
procedures and convergence results
are presented in
Appendix~\ref{app:numerical_implementation},
with the quantitative comparisons
summarized in
Table~\ref{tab:convergence_extreme}.

\section{Results and Discussion}
\label{sec:results}

\subsection{Energy deposition and numerical stability}
\label{sec:kernel_results}

The spatial characteristics of energy deposition
from the $^{10}\mathrm{B}(n,\alpha)^{7}\mathrm{Li}$
reaction products are presented in
Fig.~\ref{fig:bnct_kernel}. The radial
distribution in panel (a) shows that energy
deposition from $^{7}\mathrm{Li}$ nuclei is
more concentrated near the reaction site,
whereas alpha particles deposit energy over
longer distances. The combined energy-deposition
response generally decreases with radial
distance, reflecting the finite ranges of
these charged particles in the medium.
The absorbed dose at a given
location depends not only on the local boron
concentration but also on energy deposited
by reactions occurring in the surrounding
region. The cumulative distribution in
Fig.~\ref{fig:bnct_kernel}(b) further shows
that most of the reaction-product energy
is deposited within a few micrometers
of the capture site. At a radial resolution
of $0.05~\mu\mathrm{m}$, the radii containing
$50\%$, $90\%$, and $99\%$ of the deposited
energy are $2.49$, $6.00$, and
$7.91~\mu\mathrm{m}$, respectively.
Only approximately $0.0028\%$ of the
recorded energy is deposited beyond
$11~\mu\mathrm{m}$, indicating that the
energy-deposition response is strongly
localized around the neutron-capture site.
This spatial localization provides the
physical basis for examining how boron
concentration fluctuations are smoothed
over distances comparable to the ranges
of the reaction products.

The numerical stability of the
energy-deposition response was assessed
by comparing two Geant4 simulations,
each using $10^6$ particle histories
per source but different radial bin
widths and maximum transport step lengths.
Reducing the radial bin width from
$0.10$ to $0.05~\mu\mathrm{m}$ and
the maximum step length from $0.05$
to $0.025~\mu\mathrm{m}$ resulted in
relative changes of approximately
$0.12\%$ and $0.13\%$ in $r_{90}$
and $r_{99}$, respectively. The
characteristic energy-deposition length,
$\ell_B$, calculated from the second
moment of the spatial response,
changed by only about $0.075\%$,
with the finer simulation yielding
$\ell_B=2.048~\mu\mathrm{m}$.
These small differences indicate that
the principal spatial characteristics
of the energy-deposition response
are relatively insensitive to the
numerical settings examined. Unlike
$r_{90}$, which specifies the radius
containing a given fraction of deposited
energy, $\ell_B$ characterizes the
effective spatial width of the response
through its second moment. This
distinction is important because the
smoothing of boron-induced dose
heterogeneity depends on the full
spatial distribution of deposited
energy rather than on a single
measure of particle range. The
second-moment definition also provides
a consistent basis for comparing
the Geant4 response with a Gaussian
response of the same characteristic
length in Sec.~\ref{sec:scaling_results}.

The numerical stability of the
three-dimensional dose calculation
was examined separately from that
of the Geant4 energy-deposition
response. In particular, the grid
spacing affects the representation
of short-range boron fluctuations,
whereas the computational domain
size can influence the statistical
representation of larger spatial
structures and the effects of
periodic boundary conditions.
To evaluate these effects, additional
calculations were performed for
$H=1.2$ and a target correlation-length
ratio $\xi/\ell_B=6$, corresponding
to the largest relative difference
in $D_{95}$ within the investigated
parameter range. Grid-resolution
sensitivity was examined by increasing
the grid size from $512^3$ to
$640^3$ at a fixed domain length
of $128~\mu\mathrm{m}$, while
finite-domain sensitivity was
evaluated by increasing the domain
length from $128$ to $160~\mu\mathrm{m}$
at a fixed grid spacing of
$0.25~\mu\mathrm{m}$. Each numerical
configuration was evaluated using
20 independent random realizations,
and the resulting transfer factor
$S$ and relative difference in
$D_{95}$ were compared.

As summarized in
Table~\ref{tab:convergence_extreme}
of Appendix~\ref{app:numerical_implementation},
grid refinement changed the
ensemble-averaged transfer factor
by approximately $0.043\%$,
whereas increasing the domain
size changed it by approximately
$0.074\%$. The corresponding
differences in $\Delta_{95}$
were $0.49$ and $1.13$
percentage points, respectively.
These results indicate that
the transfer factor is relatively
stable under the numerical changes
examined, while the lower dose
percentile exhibits greater
variability. The differences in
$\Delta_{95}$ were comparable
to the statistical uncertainties
of the ensemble estimates,
reflecting the sensitivity
of the lower tail of the dose
distribution to individual
spatial realizations. Although
these comparisons do not
establish a strict bound on
finite-domain errors, they
show that the substantial
reduction in $D_{95}$ persists
under grid refinement and
domain enlargement. Together
with the stability of the
Geant4 energy-deposition response,
these findings support the
numerical robustness of the
principal results within the
tested configurations.

\subsection{Boron heterogeneity transfer to dose}
\label{sec:heterogeneity_results}

\begin{figure*}[tb!]
    \centering
    \includegraphics[width=\linewidth]
    {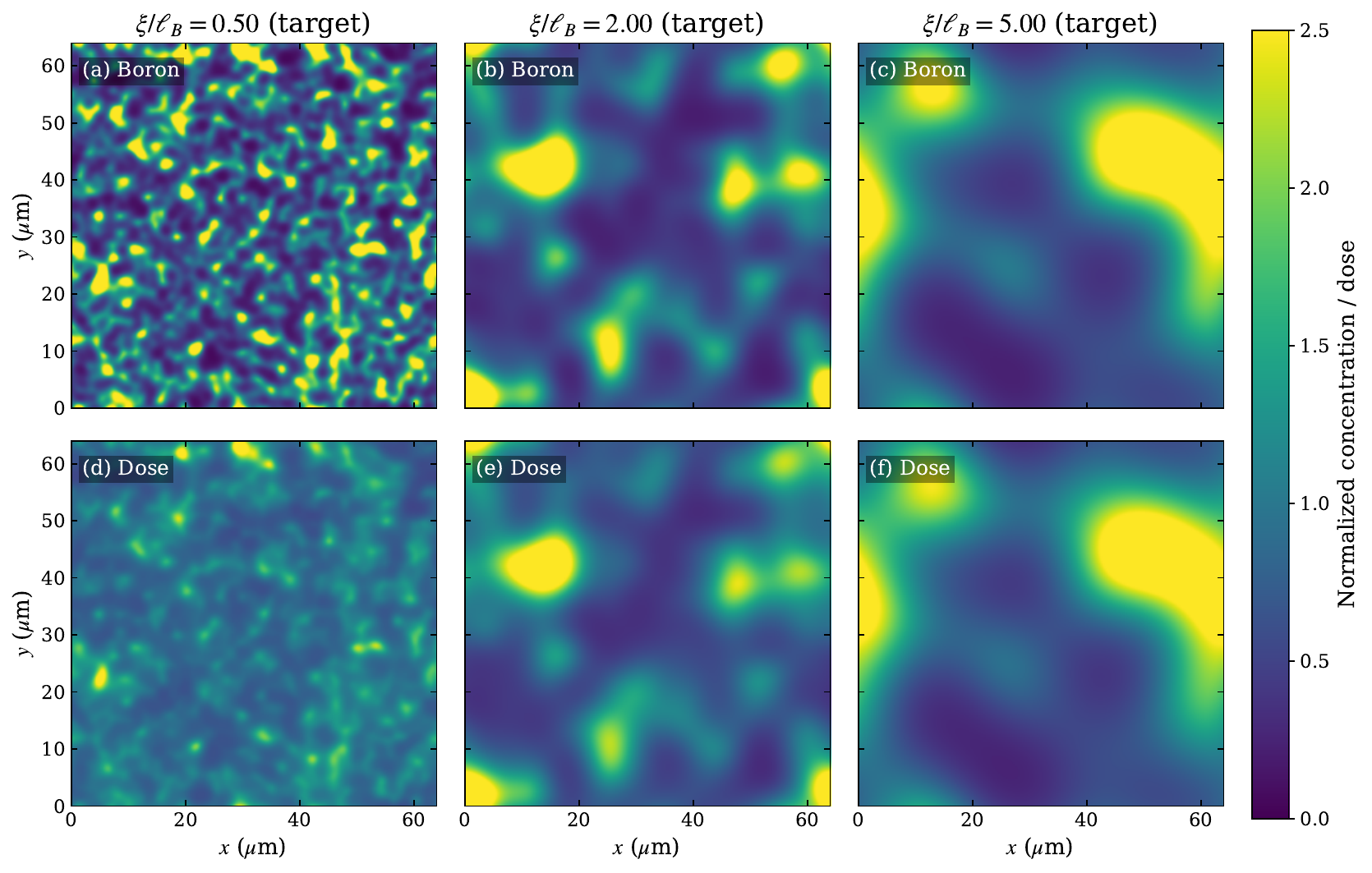}
    \caption{Comparison of spatial boron
    concentration and absorbed-dose
    distributions for three different
    correlation lengths. Panels (a)--(c)
    show the boron concentration fields,
    while panels (d)--(f) show the
    corresponding absorbed-dose distributions
    calculated by convolution with the
    Geant4 energy-deposition response.
    All configurations have the same
    boron heterogeneity amplitude, $H=0.8$,
    with target correlation-length ratios
    $\xi/\ell_B=0.5$, $2.0$, and $5.0$,
    from left to right. Calculations are
    performed on a three-dimensional
    $512^3$ grid. Concentration and dose
    are normalized by their respective
    mean values and displayed using
    a common color scale.}
    \label{fig:boron_dose_maps}
\end{figure*}

The influence of spatial boron heterogeneity
on absorbed-dose nonuniformity is first examined
by comparing the boron concentration field,
$C_B(\mathbf{r})$, with the corresponding
absorbed-dose distribution, $D_B(\mathbf{r})$.
Calculations are performed on a three-dimensional
$512^3$ grid with a domain length of
$64~\mu\mathrm{m}$ along each direction.
The mean boron concentration and heterogeneity
amplitude are held constant at $H=0.8$,
while the correlation length is varied to
obtain target ratios of $\xi/\ell_B=0.5$,
$2.0$, and $5.0$. This allows the effect
of the spatial extent of boron structures
to be examined independently of their
heterogeneity amplitude.
Figure~\ref{fig:boron_dose_maps} shows
two-dimensional slices of the boron
concentration fields and their corresponding
absorbed-dose distributions, calculated
through convolution with the Geant4
energy-deposition response using Fourier
transforms. Both concentration and dose
are normalized by their respective mean
values and displayed using a common color
scale, allowing their spatial patterns
to be compared directly.

At the shortest target correlation-length
ratio, $\xi/\ell_B=0.5$, the boron
concentration field in
Fig.~\ref{fig:boron_dose_maps}(a)
contains numerous small boron-rich and
boron-poor regions. After the reaction-product
energy spreads into the surrounding medium,
these spatial variations become considerably
smoother, as shown in
Fig.~\ref{fig:boron_dose_maps}(d).
In particular, regions with high boron
concentrations no longer produce equally
sharp dose peaks, indicating substantial
spatial smoothing when the characteristic
size of boron fluctuations is comparable
to or smaller than the energy-deposition
length. At $\xi/\ell_B=2.0$, larger
boron-rich and boron-poor structures
emerge, as illustrated in
Fig.~\ref{fig:boron_dose_maps}(b).
The corresponding dose distribution
in panel (e) still follows the locations
of boron-rich regions, although their
boundaries become smoother. Spatial
averaging therefore remains evident
but is less pronounced than at the
shorter correlation length.

At the largest target ratio,
$\xi/\ell_B=5.0$, the differences
between the boron concentration and
dose distributions become smaller.
Figures~\ref{fig:boron_dose_maps}(c)
and (f) show nearly identical spatial
patterns, indicating that the
characteristic size of the boron
structures is substantially larger
than the energy-deposition length.
In this regime, most of the reaction-product
energy is deposited within regions where
boron concentration varies relatively
little over the particle transport scale.
These comparisons demonstrate that the
correlation length strongly influences
dose nonuniformity even when the mean
boron concentration and heterogeneity
amplitude remain unchanged. Short-range
fluctuations undergo stronger smoothing,
whereas larger spatial structures are
more readily preserved in the resulting
dose distribution. Thus, dose heterogeneity
depends on both the amplitude and the
spatial scale of boron concentration
fluctuations.

\begin{figure}[pos=tb!]
    \centering
    \includegraphics[width=\linewidth]
    {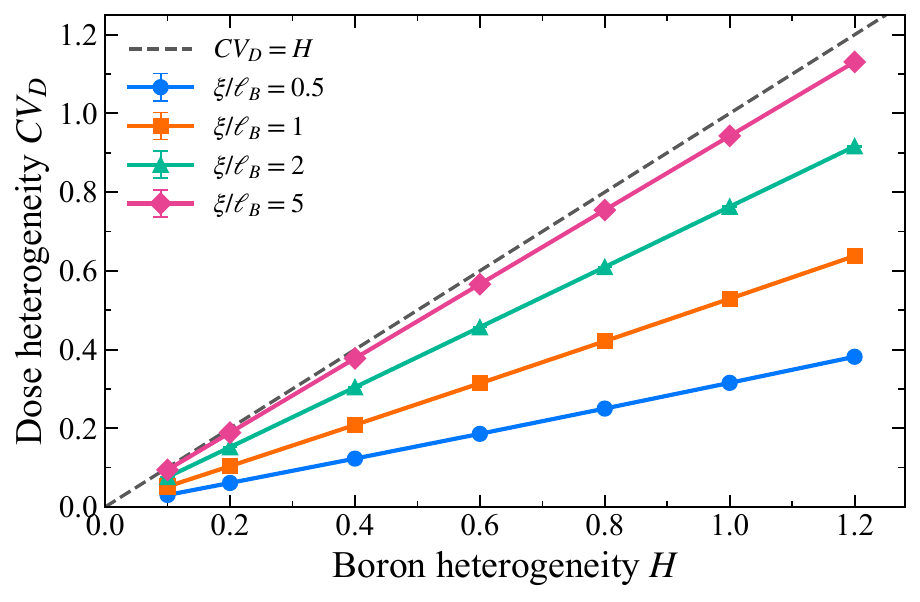}
    \caption{Coefficient of variation
    of absorbed dose, $CV_D$, as a function
    of boron concentration heterogeneity,
    $H$, for four target correlation-length
    ratios, $\xi/\ell_B=0.5$, $1$, $2$,
    and $5$. Dose distributions are
    calculated by convolving three-dimensional
    boron concentration fields with the
    Geant4 energy-deposition response.
    Symbols represent averages over
    10 independent random realizations,
    and error bars indicate the standard
    error of the mean. The dashed line
    represents the reference relation
    $CV_D=H$.}
    \label{fig:heterogeneity_transfer}
\end{figure}

To quantify this behavior, the coefficient
of variation of absorbed dose, $CV_D$,
is calculated for different heterogeneity
amplitudes, $H$, and correlation-length
ratios, $\xi/\ell_B$.
Figure~\ref{fig:heterogeneity_transfer}
shows $CV_D$ as a function of $H$
for four target correlation-length ratios,
$\xi/\ell_B=0.5$, $1$, $2$, and $5$.
The calculations use a three-dimensional
$512^3$ grid with a domain length of
$128~\mu\mathrm{m}$. Each data point
represents the average of 10 independent
random realizations, with error bars
indicating the standard error of the mean.
The dashed line $CV_D=H$ provides a
reference for the case in which boron
heterogeneity is transferred to the
absorbed-dose distribution without
attenuation.

As shown in Fig.~\ref{fig:heterogeneity_transfer},
$CV_D$ increases approximately linearly
with $H$ for all correlation lengths
considered, although the rate of increase
depends strongly on the correlation length.
At $\xi/\ell_B=0.5$, dose heterogeneity
is substantially smaller than boron
heterogeneity, with $CV_D\approx0.25$
at $H=0.8$. For the same heterogeneity
amplitude, $CV_D$ increases to approximately
$0.75$ at $\xi/\ell_B=5$, demonstrating
that larger boron structures retain
spatial dose variations more effectively.
All curves remain below the reference
line $CV_D=H$, indicating that particle
transport reduces the amplitude of spatial
fluctuations in the resulting dose
distribution. This reduction is strongest
at short correlation lengths and becomes
progressively weaker as $\xi/\ell_B$
increases. Using the definition in
Eq.~\eqref{eq:transfer_factor}, the
corresponding transfer factors at
$H\approx0.8$ are $S\approx0.31$
and $S\approx0.94$ for
$\xi/\ell_B=0.5$ and $5$, respectively.
These results show that an increasing
fraction of boron heterogeneity is
preserved in the absorbed-dose distribution
as the correlation length increases.

The approximately linear dependence
of $CV_D$ on $H$ indicates that
the heterogeneity amplitude controls
the overall magnitude of dose fluctuations,
while the different slopes demonstrate
that its influence depends on the
spatial correlation length.
This behavior is consistent with the
spatial smoothing observed in
Fig.~\ref{fig:boron_dose_maps}, where
the energy-deposition response suppresses
small-scale fluctuations more strongly
than large-scale variations.
To further characterize this relationship,
the transfer factor $S=CV_D/H$ is examined
as a function of $\xi/\ell_B$.
This representation allows us to assess
whether the transfer of heterogeneity
follows an approximate scaling relation
and to compare the numerical results
with the analytical prediction of
the ideal Gaussian model.

\subsection{Scaling and homogeneous-boron limitations}
\label{sec:scaling_results}

\begin{figure}[pos=tb!]
    \centering
    \includegraphics[width=\linewidth]
    {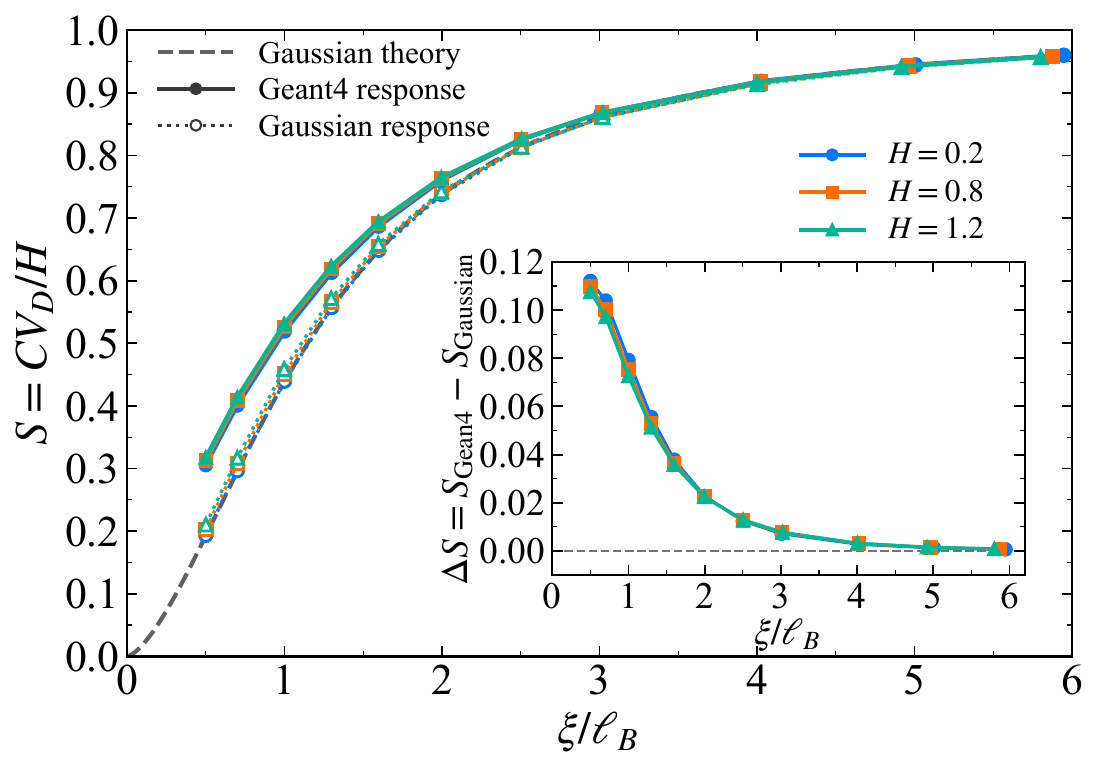}
    \caption{Scaling behavior of the
    heterogeneity transfer factor,
    $S=CV_D/H$, as a function of the
    measured correlation-length ratio
    $\xi/\ell_B$ for three boron
    heterogeneity amplitudes,
    $H=0.2$, $0.8$, and $1.2$.
    Filled symbols connected by solid
    lines represent the Geant4 response,
    while open symbols connected by dotted
    lines represent the Gaussian response
    with the same characteristic length.
    The dashed line shows the analytical
    prediction for the ideal Gaussian
    covariance and response. Both numerical
    calculations use identical lognormal
    boron fields. The inset shows the
    difference between the two numerical
    transfer factors,
    $\Delta S=S_{\mathrm{Geant4}}-
    S_{\mathrm{Gaussian}}$.}
    \label{fig:scaling_transfer}
\end{figure}

To further characterize the spatial transfer
of boron heterogeneity, the transfer factor
$S=CV_D/H$ is examined as a function of
$\xi/\ell_B$, which compares the boron
correlation length with the characteristic
energy-deposition length.
Figure~\ref{fig:scaling_transfer} presents
the numerical results for three heterogeneity
amplitudes, $H=0.2$, $0.8$, and $1.2$,
together with the analytical Gaussian
prediction in
Eq.~\eqref{eq:analytical_scaling}.
To isolate the influence of the spatial
shape of energy deposition, two response
functions are considered: the Geant4
response and a Gaussian response with
the same characteristic length $\ell_B$.
Both are applied to identical realizations
of the lognormal boron concentration
fields, allowing their effects on the
transfer factor to be compared directly.
The ideal Gaussian prediction provides
a simple reference for the dependence
on $\xi/\ell_B$, while the lognormal--Gaussian
expression in
Eq.~\eqref{eq:lognormal_gaussian_scaling}
accounts for the spatial covariance
of the boron fields used in the
numerical calculations. These two
analytical descriptions help distinguish
the influence of boron correlation
structure from that of the
energy-deposition response.

As shown in Fig.~\ref{fig:scaling_transfer},
the transfer factor increases monotonically
with $\xi/\ell_B$. At short correlation
lengths, boron concentration fluctuations
are strongly smoothed by particle transport,
resulting in relatively low values of $S$.
As the correlation length increases,
the spatial extent of boron structures
becomes larger than the energy-deposition
scale, and the absorbed-dose distribution
more closely follows the underlying boron
concentration field. This behavior is
consistent with the limiting prediction
$S\rightarrow0$ for $\xi/\ell_B\ll1$
and $S\rightarrow1$ for
$\xi/\ell_B\gg1$. Moreover, the curves
for the three heterogeneity amplitudes
nearly overlap over the investigated
range, indicating that the transfer
factor is primarily controlled by
$\xi/\ell_B$, with a relatively weak
dependence on $H$. This behavior can
also be understood from the analytical
lognormal--Gaussian model in
Eq.~\eqref{eq:lognormal_gaussian_scaling},
which retains the dependence on $H$
but predicts only modest variations
in $S$ for the parameters considered.
The approximate scaling collapse
therefore reflects the dominant role
of the correlation-length ratio within
the present model, rather than an
exact independence from heterogeneity
amplitude or a universal scaling law.

Despite their similar overall trends,
the Geant4 and Gaussian responses
produce different transfer factors
at short correlation lengths.
At $\xi/\ell_B\approx0.5$, the Geant4
response yields $S\approx0.31$,
compared with approximately $0.20$
for the numerical Gaussian response
and $0.19$ for the ideal analytical
prediction. The lognormal--Gaussian
expression provides a more appropriate
analytical reference for the numerical
Gaussian response because it retains
the covariance of the underlying
lognormal boron field. For example,
at $\xi/\ell_B=0.5$, this expression
predicts $S_{\mathrm{LG}}\approx0.19$
for $H=0.2$ and approximately $0.21$
for $H=1.2$. These values indicate
that the effect of lognormal correlation
structure is relatively small in
this regime compared with the difference
between the two energy-deposition
responses. The inset of
Fig.~\ref{fig:scaling_transfer}
shows that this difference reaches
$\Delta S\approx0.11$ at the shortest
correlation length and decreases
rapidly as $\xi/\ell_B$ increases.
Because both numerical responses are
applied to identical boron fields,
their difference can be attributed
to the spatial shape of the
energy-deposition kernel rather
than to changes in the boron
distribution. In spectral terms,
short-range boron fluctuations
contain stronger contributions
from spatial frequencies at which
the Geant4 and Gaussian responses
attenuate fluctuations differently.
Thus, matching the characteristic
energy-deposition length alone does
not guarantee equivalent smoothing,
particularly when the spatial scale
of boron heterogeneity is comparable
to or smaller than the particle
transport range.

\begin{figure}[pos=tb!]
    \centering
    \includegraphics[width=\linewidth]
    {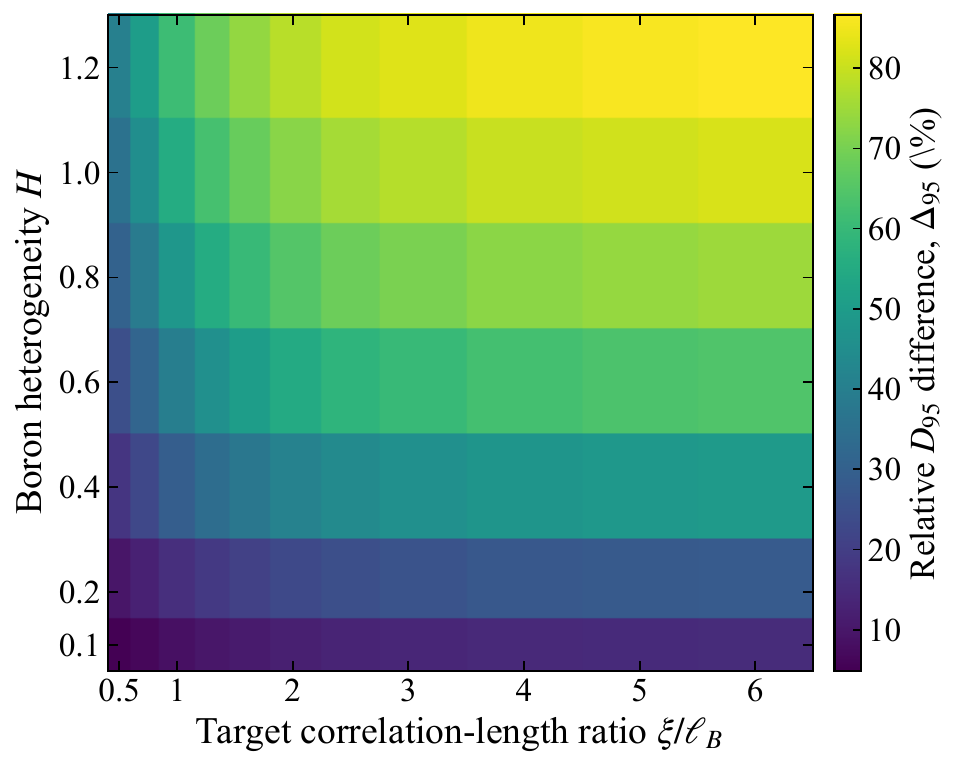}
    \caption{Relative difference in
    $D_{95}$ between homogeneous
    and heterogeneous boron
    distributions, $\Delta_{95}$,
    as a function of heterogeneity
    amplitude $H$ and the target
    correlation-length ratio
    $\xi/\ell_B$. Dose distributions
    are calculated by convolving
    three-dimensional boron
    concentration fields with the
    Geant4 energy-deposition response.
    The color scale represents the
    mean $\Delta_{95}$ over 10
    independent random realizations
    for each parameter combination.
    The largest relative difference
    is approximately $86.6\%$ at
    $H=1.2$ and $\xi/\ell_B=6$.}
    \label{fig:d95_bias}
\end{figure}

The dosimetric consequences of boron
heterogeneity are evaluated using
$D_{95}$, defined as the fifth
percentile of the absorbed-dose
distribution, corresponding to the
dose received by at least $95\%$
of the modeled volume. This quantity
is compared between homogeneous
and heterogeneous boron distributions
with the same mean concentration.
Their relative difference is
defined as
\begin{equation}
\Delta_{95}
=
\frac{
D_{95}^{\mathrm{hom}}
-
D_{95}^{\mathrm{het}}
}{
D_{95}^{\mathrm{hom}}
}
\times 100\%,
\label{eq:d95_bias}
\end{equation}
where $D_{95}^{\mathrm{hom}}$
and $D_{95}^{\mathrm{het}}$
denote the values obtained from
the homogeneous and heterogeneous
models, respectively. A positive
$\Delta_{95}$ indicates that the
homogeneous-boron approximation
overestimates $D_{95}$ relative
to the heterogeneous model.
Figure~\ref{fig:d95_bias} shows
$\Delta_{95}$ for different
heterogeneity amplitudes and
target correlation-length ratios.
The calculations use the Geant4
energy-deposition response on a
three-dimensional $512^3$ grid
with a domain length of
$128~\mu\mathrm{m}$, and each
value represents an average
over 10 independent random
realizations. Because the dose
distributions are normalized by
their respective mean values,
these comparisons isolate changes
in the spatial distribution of
the boron-dose component rather
than changes in its average value.

As shown in Fig.~\ref{fig:d95_bias},
the relative difference in $D_{95}$
increases with both heterogeneity
amplitude and correlation length.
At a target ratio $\xi/\ell_B=2$,
$\Delta_{95}$ increases from
approximately $22.8\%$ at $H=0.2$
to $64.6\%$ at $H=0.8$ and
$77.7\%$ at $H=1.2$. The
correlation-length dependence
is also pronounced: at $H=0.8$,
$\Delta_{95}$ increases from
approximately $30.4\%$ at
$\xi/\ell_B=0.5$ to $74.5\%$
at $\xi/\ell_B=6$. At short
correlation lengths, particle
transport smooths differences
between boron-rich and boron-poor
regions, reducing the spatial
variation in absorbed dose.
As the correlation length increases,
however, boron-poor regions extend
over larger distances relative
to the energy-deposition range
and can retain lower absorbed doses.
This behavior reduces the lower
percentiles of the dose distribution,
even when the mean dose remains
unchanged. The largest difference
within the investigated parameter
range occurs at $H=1.2$ and a
target ratio $\xi/\ell_B=6$, where
$\Delta_{95}=(86.62\pm1.07)\%$,
with the uncertainty expressed
as the standard error in percentage
points. Under these conditions,
$D_{95}$ in the heterogeneous model
is only approximately $13.4\%$
of the corresponding homogeneous
value.

The behavior at large correlation
lengths can also be understood
from the underlying lognormal
distribution. When $\xi$ becomes
much larger than the energy-deposition
length, spatial smoothing becomes
weak and the normalized dose
distribution approaches the
normalized boron concentration
distribution. In this limit,
the fifth percentile follows
directly from the lognormal model
in Eq.~\eqref{eq:boron_field},
giving
\begin{equation}
\Delta_{95}^{(\infty)}
=
\left[
1-
\exp\left(
-\frac{s^2}{2}
+s\Phi^{-1}(0.05)
\right)
\right]\times100\%,
\label{eq:d95_lognormal_limit}
\end{equation}
where $\Phi^{-1}$ is the inverse
cumulative distribution function
of the standard normal distribution.
For $H=1.2$, this expression
predicts $\Delta_{95}^{(\infty)}
\approx86.5\%$, which is close
to the largest numerical value
obtained in Fig.~\ref{fig:d95_bias}.
The agreement is consistent with
the expected reduction of spatial
smoothing at large correlation
lengths, although it does not
establish that the asymptotic
limit has been fully reached
within the simulated domain.
This analytical limit also
provides a useful reference
for interpreting the magnitude
of the changes in $D_{95}$
without introducing additional
assumptions about the biological
response to radiation.

The results in
Fig.~\ref{fig:d95_bias} complement
the scaling behavior in
Fig.~\ref{fig:scaling_transfer}.
Although the transfer factor $S$
depends only weakly on $H$ over
the investigated range, the
relative difference in $D_{95}$
remains strongly dependent on
heterogeneity amplitude. This
distinction arises because $S$
characterizes the overall amplitude
of dose fluctuations, whereas
$D_{95}$ is determined by the
lower tail of the dose distribution.
Similar transfer
factors do not necessarily imply
equivalent dose coverage, particularly
when the underlying concentration
distributions differ in their
fluctuation amplitude. More generally,
these findings demonstrate that
the mean boron concentration alone
is insufficient to characterize
absorbed-dose nonuniformity in
a heterogeneous medium. Both the
amplitude and spatial correlation
length of boron fluctuations,
together with the shape of the
energy-deposition response,
contribute to the resulting dose
distribution. Because the present
model considers only the boron-dose
component in a homogeneous medium
under a uniform neutron field,
the calculated differences in
$D_{95}$ should not be interpreted
as differences in total clinical
BNCT dose or biological effectiveness.
Instead, they provide a physical
basis for assessing the limitations
of the homogeneous-boron approximation
in spatial dosimetric analysis.

\section{Conclusions}
\label{sec:conclusions}

This study investigated the influence of
boron heterogeneity amplitude and spatial
correlation length on absorbed-dose
nonuniformity in boron neutron capture
therapy (BNCT). By combining Geant4
particle transport simulations, correlated
lognormal boron concentration fields,
and three-dimensional convolution, we
examined how the spatial structure of
boron concentration affects the resulting
absorbed-dose distribution. The Geant4
simulations showed that approximately
$90\%$ of the reaction-product energy
is deposited within a radius of
$6.00~\mu\mathrm{m}$, with a characteristic
energy-deposition length of
$\ell_B\approx2.05~\mu\mathrm{m}$.
This length provides a characteristic
scale for understanding how charged-particle
transport smooths boron concentration
variations at the micrometer scale.

The numerical results demonstrated that
dose heterogeneity depends on both the
amplitude and spatial correlation length
of boron concentration fluctuations.
Small-scale fluctuations undergo stronger
smoothing, whereas larger boron structures
are more readily preserved in the
absorbed-dose distribution. The
heterogeneity transfer factor,
$S=CV_D/H$, exhibited an approximate
scaling collapse as a function of
$\xi/\ell_B$, with relatively weak
dependence on heterogeneity amplitude
over the investigated parameter range.
However, comparison with the Gaussian
reference model revealed that the
characteristic energy-deposition length
alone does not fully determine the
transfer of boron heterogeneity.
At $\xi/\ell_B\approx0.5$, the Geant4
response yielded $S\approx0.31$,
compared with approximately $0.20$
for a Gaussian response with the same
characteristic length. The difference
decreased as the correlation length
increased, highlighting the importance
of the spatial shape of energy deposition,
particularly at short correlation lengths.

The dosimetric consequences were further
examined using $D_{95}$, whose relative
difference between homogeneous and
heterogeneous boron models increased
with both heterogeneity amplitude and
correlation length. At $H=1.2$ and a
target correlation-length ratio of
$\xi/\ell_B=6$, the relative difference
reached $\Delta_{95}=(86.62\pm1.07)\%$,
where the uncertainty represents the
standard error in percentage points.
This result demonstrates that assuming
a uniform boron concentration can
substantially overestimate $D_{95}$
when the underlying distribution is
heterogeneous. Nevertheless, because
the present model considers only the
boron-dose component in a homogeneous
medium under a uniform neutron field,
the calculated differences cannot be
directly interpreted as changes in
total clinical BNCT dose. Our results
establish a physical basis for assessing
the limitations of the homogeneous-boron
approximation and emphasize the combined
roles of heterogeneity amplitude,
spatial correlation length, and
energy-deposition response shape
in determining absorbed-dose
nonuniformity.

\section*{Data Availability}
\textcolor{black}{The simulation data and source codes supporting the results of this study are available from the corresponding author upon reasonable request.}

\section*{CRediT authorship contribution statement}

\textbf{R.~Muslim:}
Conceptualization, Methodology,
\textcolor{black}{Writing -- original draft},
Software, Formal analysis, Validation,
\textcolor{black}{Visualization},
\textcolor{black}{Writing -- review \& editing}. \textbf{B.~Pidhatika:}
Conceptualization, Formal analysis,
\textcolor{black}{Writing -- original draft},
Validation,
\textcolor{black}{Writing -- review \& editing}. \textbf{S.~Sutanto:}
Formal analysis, Validation,
\textcolor{black}{Writing -- review \& editing}. \textbf{R.~Fardela:}
Formal analysis, Validation,
\textcolor{black}{Writing -- review \& editing}. \textbf{R.~N.~Wijaya:}
Formal analysis, Validation,
\textcolor{black}{Writing -- review \& editing}. 
\textcolor{black}{Writing -- review \& editing}. All authors have read, reviewed, and approved the publication of this paper.

\section*{Declaration of Interests}
The authors declare that they have no known competing financial interests or personal relationships that could have appeared to influence the work reported in this paper.


\section*{Acknowledgments}
\textbf{R.~Muslim} acknowledges support from the YST Program of the Asia Pacific Center for Theoretical Physics (APCTP), funded by the Science and Technology Promotion Fund and Lottery Fund of the Korean Government, and by the Management Talent Program of the National Research and Innovation Agency of Indonesia (BRIN).

\printcredits

\appendix

\section{Gaussian and lognormal scaling theory}
\label{app:gaussian_scaling}
In this section, we derive the main
analytical expressions presented in
the main text. Starting from the
convolution model in
Eq.~\eqref{eq:dose_convolution},
the normalized energy-deposition
response satisfies
$\int K_B(\mathbf r)\,d^3\mathbf r=1$,
which gives
$\langle D_B\rangle=A\bar C_B$.
The Fourier transforms of the
fluctuating components are related by
\begin{equation}
\widetilde{\delta D}_B(\mathbf k)
=
A\widetilde K_B(\mathbf k)
\widetilde{\delta C}_B(\mathbf k).
\label{eq:app_fourier}
\end{equation}
The corresponding variances are
\begin{align}
\sigma_D^2
&=
A^2\int
\left|\widetilde K_B(\mathbf k)\right|^2
S_B(\mathbf k)
\frac{d^3\mathbf k}{(2\pi)^3},
\label{eq:app_dose_variance}\\
\sigma_B^2
&=
\int S_B(\mathbf k)
\frac{d^3\mathbf k}{(2\pi)^3}.
\label{eq:app_boron_variance}
\end{align}
Using $S=CV_D/H$ and
$\langle D_B\rangle=A\bar C_B$,
we obtain
\begin{equation}
S^2
=
\frac{\sigma_D^2}{A^2\sigma_B^2}
=
\frac{
\int
S_B(\mathbf k)
|\widetilde K_B(\mathbf k)|^2
\,d^3\mathbf k
}{\int
S_B(\mathbf k)\,d^3\mathbf k
},
\label{eq:app_transfer_spectrum}
\end{equation}
which reproduces
Eq.~\eqref{eq:transfer_spectrum}.

For the Gaussian reference model,
the boron concentration covariance is
\begin{equation}
G_B(r)
=
\sigma_B^2
\exp\left(-\frac{r^2}{2\xi^2}\right).
\label{eq:app_gaussian_covariance}
\end{equation}
Its three-dimensional Fourier
transform gives
\begin{equation}
S_B(k)
=
(2\pi)^{3/2}\sigma_B^2\xi^3
\exp\left(-\frac{\xi^2k^2}{2}\right).
\label{eq:app_gaussian_spectrum}
\end{equation}
The normalized Gaussian
energy-deposition response is
\begin{equation}
K_G(r)=
\frac{1}{(2\pi\ell_B^2)^{3/2}}
\exp\left(-\frac{r^2}{2\ell_B^2}\right),
\label{eq:app_gaussian_kernel}
\end{equation}
where
$\ell_B^2=\langle r^2\rangle_K/3$.
Its Fourier transform is
\begin{equation}
\widetilde K_G(k)
=
\exp\left(-\frac{\ell_B^2k^2}{2}\right).
\label{eq:app_gaussian_kernel_fourier}
\end{equation}
Substituting Eq.~\eqref{eq:app_gaussian_spectrum} and
Eq.~\eqref{eq:app_gaussian_kernel_fourier}
into Eq.~\eqref{eq:app_transfer_spectrum}
yields
\begin{align}
S_G^2
&=
\frac{
\displaystyle\int_{\mathbb R^3}
e^{-(\xi^2/2+\ell_B^2)k^2}
\,d^3\mathbf k
}{
\displaystyle\int_{\mathbb R^3}
e^{-\xi^2k^2/2}
\,d^3\mathbf k
}=
\frac{
\left[\pi/(\xi^2/2+\ell_B^2)\right]^{3/2}
}{
\left(2\pi/\xi^2\right)^{3/2}
}
\nonumber\\
&=
\left(
\frac{\xi^2}
{\xi^2+2\ell_B^2}
\right)^{3/2},
\label{eq:app_gaussian_integral}
\end{align}
where
$\int_{\mathbb R^3}
e^{-bk^2}\,d^3\mathbf k
=(\pi/b)^{3/2}$.
Taking the positive square root gives
\begin{equation}
S_G=
\left(
\frac{\xi^2}
{\xi^2+2\ell_B^2}
\right)^{3/4},
\label{eq:app_scaling_result}
\end{equation}
in agreement with
Eq.~\eqref{eq:analytical_scaling}.

We consider the correlated
lognormal field defined in
Eq.~\eqref{eq:boron_field}.
Let $X_1=X(\mathbf r)$,
$X_2=X(\mathbf r')$, and
\begin{equation}
\rho_X(r)
=
\langle X_1X_2\rangle
=
\exp\left(-\frac{r^2}{2a^2}\right).
\end{equation}
Since $X_1$ and $X_2$ are
jointly Gaussian with zero mean
and unit variance,
\begin{align}
\left\langle e^{s(X_1+X_2)}\right\rangle
&=
\exp\left[
\frac{s^2}{2}
\operatorname{Var}(X_1+X_2)
\right]
\nonumber\\
&=
\exp\left[
s^2(1+\rho_X(r))
\right],
\label{eq:app_joint_gaussian_moment}
\end{align}
where
$\operatorname{Var}(X_1+X_2)
=2[1+\rho_X(r)]$.
It follows that
\begin{align}
\left\langle
C_B(\mathbf r)C_B(\mathbf r')
\right\rangle
&=
\bar C_B^2e^{-s^2}
\left\langle e^{s(X_1+X_2)}\right\rangle
\nonumber\\
&=
\bar C_B^2
\exp\left[s^2\rho_X(r)\right].
\end{align}
Subtracting the squared mean
concentration gives
\begin{align}
G_B(r)
&=
\bar C_B^2
\left[
\exp\left(
s^2e^{-r^2/(2a^2)}
\right)-1
\right].
\label{eq:app_lognormal_covariance}
\end{align}
Expanding the exponential, we obtain
\begin{equation}
G_B(r)
=
\bar C_B^2
\sum_{m=1}^{\infty}
\frac{s^{2m}}{m!}
\exp\left(
-\frac{mr^2}{2a^2}
\right).
\label{eq:app_lognormal_expansion}
\end{equation}

Using the Fourier transform identity
\begin{equation}
\int_{\mathbb R^3}
e^{-r^2/(2b^2)}
e^{-i\mathbf k\cdot\mathbf r}
\,d^3\mathbf r
=
(2\pi)^{3/2}b^3
e^{-b^2k^2/2},
\end{equation}
with $b=a/\sqrt{m}$,
Eq.~\eqref{eq:app_lognormal_expansion}
gives
\begin{equation}
\begin{aligned}
S_B^{\mathrm{LG}}(k)
={}&
(2\pi)^{3/2}\bar C_B^2a^3
\sum_{m=1}^{\infty}
\frac{s^{2m}}{m!\,m^{3/2}}
\exp\left(
-\frac{a^2k^2}{2m}
\right).
\end{aligned}
\label{eq:app_lognormal_spectrum}
\end{equation}

For the Gaussian response,
$|\widetilde K_G(k)|^2
=\exp(-\ell_B^2k^2)$.
The contribution of each term
to the normalized dose variance
involves the integral
\begin{align}
I_m
&={}
\frac{(2\pi)^{3/2}a^3}{m^{3/2}}
\int_{\mathbb R^3}
\exp\left[
-\left(
\frac{a^2}{2m}+\ell_B^2
\right)k^2
\right]
\frac{d^3\mathbf k}{(2\pi)^3} \nonumber \\
&=
\frac{a^3}{m^{3/2}}
\left(
\frac{a^2}{m}+2\ell_B^2
\right)^{-3/2}
\nonumber\\
&=
\left(
\frac{a^2}
{a^2+2m\ell_B^2}
\right)^{3/2}.
\label{eq:app_lognormal_integral}
\end{align}
Substituting
Eq.~\eqref{eq:app_lognormal_spectrum}
and
Eq.~\eqref{eq:app_gaussian_kernel_fourier}
into Eq.~\eqref{eq:app_dose_variance},
and using
Eq.~\eqref{eq:app_lognormal_integral},
we obtain
\begin{equation}
\frac{\sigma_D^2}{A^2}
=
\bar C_B^2
\sum_{m=1}^{\infty}
\frac{s^{2m}}{m!}
\left(
\frac{a^2}
{a^2+2m\ell_B^2}
\right)^{3/2}.
\label{eq:app_lognormal_dose_variance}
\end{equation}
At $r=0$, the boron variance
follows directly from
Eq.~\eqref{eq:app_lognormal_covariance},
\begin{equation}
\sigma_B^2
=
G_B(0)
=
\bar C_B^2(e^{s^2}-1)
=
\bar C_B^2H^2.
\label{eq:app_lognormal_boron_variance}
\end{equation}
Therefore,
\begin{equation}
S_{\mathrm{LG}}^2
=
\frac{1}{e^{s^2}-1}
\sum_{m=1}^{\infty}
\frac{s^{2m}}{m!}
\left(
\frac{a^2}
{a^2+2m\ell_B^2}
\right)^{3/2},
\label{eq:app_lognormal_scaling}
\end{equation}
which reproduces
Eq.~\eqref{eq:lognormal_gaussian_scaling}.

The Gaussian reference
result is recovered in the
weak-heterogeneity limit.
For $H\to0^+$,
\begin{equation}
s^2=\ln(1+H^2)\sim H^2,
\qquad
a\to\xi.
\end{equation}
Keeping the leading term
$m=1$ in
Eq.~\eqref{eq:app_lognormal_scaling},
\begin{align}
S_{\mathrm{LG}}^2
&\sim
\frac{s^2}{e^{s^2}-1}
\left(
\frac{a^2}
{a^2+2\ell_B^2}
\right)^{3/2}
\nonumber\\
&\longrightarrow
\left(
\frac{\xi^2}
{\xi^2+2\ell_B^2}
\right)^{3/2}.
\end{align}
Thus,
\begin{equation}
\lim_{H\to0^+}S_{\mathrm{LG}}
=
\left(
\frac{\xi^2}
{\xi^2+2\ell_B^2}
\right)^{3/4}
=
S_G.
\label{eq:app_lognormal_gaussian_limit}
\end{equation}
Both models also satisfy
$S\to0$ for $\xi/\ell_B\to0$
and $S\to1$ for
$\xi/\ell_B\to\infty$,
under their respective
continuum assumptions.

\section{Numerical implementation}
\label{app:numerical_implementation}

The numerical procedure consists of
two stages: obtaining the spatial
energy-deposition response from Geant4
simulations and calculating absorbed-dose
distributions by convolving this response
with heterogeneous boron concentration
fields. This separation allows different
boron configurations to be examined
without repeating particle transport
simulations and enables the effects
of boron heterogeneity and energy
deposition to be investigated independently.

Alpha particles and lithium nuclei
are transported in water using $10^6$
particle histories for each particle
species and reaction branch. Deposited
energy is scored in concentric spherical
shells with a radial bin width of
$\Delta r=0.05~\mu\mathrm{m}$ and a
maximum transport step length of
$0.025~\mu\mathrm{m}$. Contributions
from the reaction branches are combined
according to their branching probabilities
to construct the total energy-deposition
response $K_B(r)$. The energy deposited
in each radial shell, $E_j$, is normalized
as $w_j=E_j/\sum_m E_m$, satisfying
$\sum_j w_j=1$. These weights describe
the relative spatial distribution of
deposited energy independently of
the total reaction-product energy.

For an isotropic response, the
three-dimensional Fourier transform
depends only on $k=|\mathbf{k}|$,
\begin{equation}
\widetilde K_B(k)
=
4\pi\int_0^\infty
r^2K_B(r)
\frac{\sin(kr)}{kr}\,dr.
\label{eq:app_radial_transform}
\end{equation}
The integral is evaluated discretely
using the normalized radial energy
weights. The condition
$\widetilde K_B(0)=1$ ensures that
convolution preserves the mean
boron-dose component. The characteristic
energy-deposition length is obtained
from the second moment,
$\ell_B^2=\langle r^2\rangle_K/3$,
providing a common spatial scale
for comparing the Geant4 and
Gaussian responses.

Boron concentration fields are generated
on a three-dimensional periodic grid.
A correlated Gaussian field $X(\mathbf r)$
is constructed by filtering Gaussian
random noise in Fourier space using
the correlation spectrum corresponding
to Eq.~\eqref{eq:spatial_correlation},
consistent with spectral approaches
for simulating stationary Gaussian
random fields \cite{wood1994simulation}.
The field is normalized to zero mean
and unit variance before applying
the lognormal transformation in
Eq.~\eqref{eq:boron_field}.
For prescribed heterogeneity amplitude
$H$ and target correlation length $\xi$,
the parameters $s$ and $a$ are determined
from the lognormal model and the
correlation-length mapping in
Sec.~\ref{sec:boron_distribution}.
The lognormal parameter is adjusted
to reproduce the prescribed
heterogeneity amplitude, and each
boron field is normalized by its
spatial mean. The actual correlation
length is measured from the normalized
spatial autocorrelation function at
its $\exp(-1/2)$ crossing. Comparing
the measured and target values allows
finite-domain fluctuations and
discretization effects to be assessed.

The absorbed-dose distribution is
calculated by Fourier-space convolution,
\begin{equation}
D_B(\mathbf r)
=
A\mathcal F^{-1}
\left[
\widetilde K_B(|\mathbf k|)
\mathcal F\{C_B(\mathbf r)\}
\right],
\label{eq:app_fft_convolution}
\end{equation}
where $\mathcal F$ and $\mathcal F^{-1}$
denote the forward and inverse
Fourier transforms. The convolution
is evaluated using three-dimensional
fast Fourier transforms implemented
with the FFTW3 library
\cite{frigo2005fftw3}, under periodic
boundary conditions. Since dose
distributions are normalized by their
spatial means, the factor $A$ does
not affect $CV_D$, the transfer
factor $S$, or the relative difference
in $D_{95}$. The Geant4 and Gaussian
responses are applied to identical
boron realizations to isolate the
influence of the energy-deposition
kernel shape.

The spatial visualizations in
Fig.~\ref{fig:boron_dose_maps}
use a $512^3$ grid with a domain
length of $64~\mu\mathrm{m}$,
corresponding to a grid spacing
of $0.125~\mu\mathrm{m}$.
The quantitative calculations
in Figs.~\ref{fig:heterogeneity_transfer},
\ref{fig:scaling_transfer}, and
\ref{fig:d95_bias} use a $512^3$
grid with a domain length of
$128~\mu\mathrm{m}$ and a spacing
of $0.250~\mu\mathrm{m}$.
For each parameter combination,
results are averaged over $M=10$
independent random realizations.
Statistical uncertainties are
reported as standard errors
of the mean,
\begin{equation}
\mathrm{SEM}(Y)
=
\frac{\mathrm{SD}(Y)}{\sqrt{M}},
\label{eq:app_standard_error}
\end{equation}
where $Y$ denotes the observable
evaluated for each realization.
The dose-coverage metric $D_{95}$
is defined as the fifth percentile
of the dose distribution normalized
by its mean. Since the normalized
dose equals unity for homogeneous
boron concentration, the relative
difference between homogeneous
and heterogeneous models is
\begin{equation}
\Delta_{95}
=
100\left(1-D_{95}^{\mathrm{het}}\right)\%,
\label{eq:app_d95_normalized}
\end{equation}
where $D_{95}^{\mathrm{het}}$
is the fifth percentile of the
normalized heterogeneous dose.
All comparisons therefore use
the same mean boron concentration.

Numerical sensitivity was assessed
through grid refinement at a fixed
domain size and domain enlargement
at a fixed grid spacing. The tests
focused on $H=1.2$ and a target
ratio $\xi/\ell_B=6$, corresponding
to the largest relative difference
in $D_{95}$ within the investigated
range. This long-correlation case
also probes the sensitivity of
low-dose regions to finite-domain
effects. Each configuration was
evaluated using 20 independent
realizations, with the transfer
factor $S$ and relative difference
$\Delta_{95}$ used to quantify
numerical sensitivity.

\begin{table}[pos=tb!]
\centering
\caption{Numerical sensitivity of the
heterogeneity transfer factor $S$
and relative difference $\Delta_{95}$
for $H=1.2$ and target
$\xi/\ell_B=6$. Results are
ensemble averages over 20 independent
realizations. Parentheses indicate
standard errors in the last
reported digits.}
\label{tab:convergence_extreme}
\small
\setlength{\tabcolsep}{3pt}
\begin{tabular*}{\columnwidth}
{@{}c@{\extracolsep{\fill}}cccc@{}}
\hline
$N$ &
$L$ ($\mu$m) &
$\Delta x$ ($\mu$m) &
$S$ &
$\Delta_{95}$ (\%) \\
\hline
512 & 128 & 0.25 &
$0.95691(77)$ & $87.08(71)$ \\
640 & 128 & 0.20 &
$0.95649(91)$ & $86.59(56)$ \\
640 & 160 & 0.25 &
$0.95762(68)$ & $85.95(56)$ \\
\hline
\end{tabular*}
\end{table}

As shown in
Table~\ref{tab:convergence_extreme},
refining the grid from $512^3$
to $640^3$ at $L=128~\mu\mathrm{m}$
changed the ensemble-averaged $S$
by approximately $0.043\%$ and
$\Delta_{95}$ by $0.49$ percentage
points. Increasing the domain length
from $128$ to $160~\mu\mathrm{m}$
at a fixed spacing of
$0.25~\mu\mathrm{m}$ changed $S$
by $0.074\%$ and $\Delta_{95}$
by $1.13$ percentage points.
The corresponding measured
correlation lengths were
$11.874$, $11.832$, and
$12.002~\mu\mathrm{m}$ for
the three configurations,
respectively, compared with
the target value of
$12.289~\mu\mathrm{m}$.
These deviations and the use
of independent random realizations
introduce ensemble variability
into the numerical comparisons.
Nevertheless, $S$ remained
relatively stable, while the
larger variations in $\Delta_{95}$
were comparable to its statistical
uncertainties. The substantial
reduction in $D_{95}$ persisted
under both grid refinement and
domain enlargement, supporting
the numerical robustness of
the principal findings within
the tested configurations,
although a strict bound on
finite-domain errors in
$D_{95}$ cannot be established.

\bibliographystyle{elsarticle-num}

\bibliography{cas-refs}







\end{document}